\documentclass{ceurart} 
\usepackage{graphicx} 
\usepackage[T1]{fontenc}
\usepackage{amsmath}
\usepackage[english]{babel}
\usepackage{array,tabularx}
\usepackage{booktabs} 
\usepackage{makeidx}  
\usepackage{xparse}
\usepackage{moredefs}
\usepackage{lips}
\usepackage{epstopdf} 
\usepackage[frozencache,cachedir=.]{minted}
\usepackage{color}
\usepackage{csquotes}
\usepackage{xcolor}
\usepackage{hyperref}
\usepackage{comment}
\usepackage{paralist}
\usepackage{cleveref}
\usepackage{wrapfig}
\usepackage{multirow}
\usepackage{microtype}
\usepackage{listings}
\usepackage[super]{nth}
\usepackage{natbib}
\usepackage{paralist}

\usepackage{float}
\floatstyle{plaintop}
\restylefloat{table}

\usepackage[style=base,labelfont=bf,labelsep=period,tableposition=top,font=small]{caption}
\makeatletter
\renewcommand\@biblabel[1]{#1.}
\makeatother
\addto\captionsenglish{}

\crefformat{footnote}{#2\footnotemark[#1]#3}
\expandafter\def\expandafter\UrlBreaks\expandafter{\UrlBreaks
  \do\a\do\b\do\c\do\d\do\e\do\f\do\g\do\h\do\i\do\j%
  \do\k\do\l\do\m\do\n\do\o\do\p\do\q\do\r\do\s\do\t%
  \do\u\do\v\do\w\do\x\do\y\do\z\do\A\do\B\do\C\do\D%
  \do\E\do\F\do\G\do\H\do\I\do\J\do\K\do\L\do\M\do\N%
  \do\O\do\P\do\Q\do\R\do\S\do\T\do\U\do\V\do\W\do\X%
  \do\Y\do\Z}

\newcolumntype{L}[1]{>{\raggedright\arraybackslash}p{#1}}   
\newcolumntype{C}[1]{>{\centering\arraybackslash}p{#1}}     
\newcolumntype{R}[1]{>{\raggedleft\arraybackslash}p{#1}}    

\begin{document}

\copyrightyear{2023}
\copyrightclause{Copyright for this paper by its authors. Use permitted under Creative Commons License Attribution 4.0 International (CC BY 4.0).} 
\conference{ITADATA2023: The 2$^{\text{nd}}$ Italian Conference on Big Data and Data Science, September 11--13, 2023, Naples, Italy}



\title{A Survey of Dataspace Connector Implementations}

\author{Tobias Dam}[%
orcid=0000-0002-2463-5831,
email=tobias.dam@fhstp.ac.at
]
\address{St.\ Pölten University of Applied Sciences, Austria}

\author{Lukas Daniel Klausner}[%
orcid=0000-0003-3650-9733,
email=mail@l17r.eu
]

\author{Sebastian Neumaier}[%
orcid=0000-0002-9804-4882,
email=sebastian.neumaier@fhstp.ac.at
]

\author{Torsten Priebe}[%
orcid=0000-0001-9282-2535,
email=torsten.priebe@fhstp.ac.at
]




\begin{abstract}
The concept of dataspaces aims to facilitate secure and sovereign data exchange among multiple stakeholders. Technical implementations known as ``connectors'' support the definition of usage control policies and the verifiable enforcement of such policies. 
This paper provides an overview of existing literature and reviews current open-source dataspace connector implementations that are compliant with the International Data Spaces (IDS) standard.
To assess maturity and readiness, we review four implementations with regard to their architecture, underlying data model and usage control language.

\end{abstract}

\maketitle




\section{Introduction} \label{sec:intro}

Dataspaces
are a concept that is gaining attention from industry and research communities worldwide. It serves as an abstraction for data management in situations where multiple stakeholders exchange data with each other. The idea is that the easy exchange of data between stakeholders generates value, particularly when combined with data analytics. 
New trading mechanisms are meant to enable stakeholders to cooperate with each other based on the value of the exchanged data and analytics services. For example, in a smart city scenario, a public transportation company and local businesses might participate in a dataspace where businesses benefit from improved retail demand predictions and the transportation company can optimize traffic management. 
This, however, requires a data management architecture which allows sharing the participants' data under well-defined and strictly controlled usage policies. 

Therefore, the goal of dataspaces is to allow members to share data offers and transfer data while keeping control over its use. Control in the context of dataspaces relates to four requirements: (i) participants remains in sovereign control of their \textit{identity}; (ii) participants decide who to \textit{trust}; (iii) participants decide on the usage \textit{policies} under which the data is shared; (iv) participants remain in \textit{control} of their deployment.

Within the overall architecture of a dataspace (cf. \Cref{fig:connectorarchitecture}), the technical implementations of these requirements are called ``connectors'', trustworthy software components that support the definition of usage control policies and the verifiable enforcement of such policies. The idea is that these connectors provide the capabilities to automate the connection, the contract negotiation, and the policy enforcement of data transfers in dataspaces. They therefore act as logical gatekeepers that integrate into each participant’s infrastructure and communicate with each other. While there are also additional services and components that are necessary for a dataspace ecosystem (such as the identity provider), the connectors are intended to be the central technical component.

Efforts to build dataspaces are currently led by Gaia-X~\citep{eggers2020gaia} and the International Data Spaces Association (IDSA)~\citep{Bader2020TheContent}; the synergies between these two are receiving broad attention in Europe and beyond. Furthermore, there are other initiatives currently working on developing the necessary building blocks and infrastructure. With the proliferation of dataspaces initiatives and the increase in both their memberships and implementations, it has become increasingly difficult to follow the underlying technical developments of these systems. The goal of this paper is to highlight the two main challenges that the dataspaces initiatives currently face: (i) the lack of a common definitions for dataspaces and their core components (the connectors), which can impede interoperability, 
and (ii) the potential lack of reusable core components, which could lead to divergent implementations and waste of effort.

In this article, we provide a study of existing developments and implementations for building dataspaces. Our work is motivated by a general, increasing interest in these technologies which is, as indicated above, also foreseeable to prevail for the next few years. 
In summary, the key contributions of our work are as follows:
\begin{compactitem}
    \item We provide an overview of the existing literature on the origins, existing definitions and technical foundations of dataspaces. 
    \item We survey existing open-source implementations of dataspace connectors with regard to their architecture, underlying data model and usage control support. 
    In this survey we focus on IDS-compliant implementations.
    \item We point out the surprisingly bad state of documentation for major open-source dataspace connectors, which runs the risk of being a major hindrance to uptake, further development and collaboration/interconnection between different technologies.
\end{compactitem}

The remainder of this paper is organized as follows: We review existing literature in \Cref{sec:background}; the survey of existing open-source dataspace connectors is in \Cref{sec:survey}. Finally, we conclude with our main results and an outlook on future work in \Cref{sec:conclusion}.

\section{Background and Related Work\label{sec:background}}

The initial and most significant development was the foundation of the International Data Spaces Association (IDSA), originally launched by the Fraunhofer Society in 2015.
The non-profit association has developed a reference architecture~\citep{idsram} and an information model~\citep{ Bader2020TheContent}, and provides tools and resources for implementing dataspaces (including a connector discussed in \Cref{ssec:dsc}). The work of the IDSA has been published in a number of articles~\citep{Otto2019DesigningCase, Zrenner2019UsageEcosystems,Bader2020TheContent,DBLP:conf/isola/PampusJQ22} and in a comprehensive open-access book~\citep{2022DesigningSpaces}.

Another notable development in the field of dataspaces is the Eclipse Dataspace Connector (EDC) project. The EDC is being developed and published as an open-source solution at the Eclipse Foundation. The goals of the EDC initiative include a concrete implementation of the IDS standard protocols and alignment of its code with the requirements of the Gaia-X project.\footnote{\url{https://github.com/eclipse-edc/Collateral/tree/main/Latest\%20Presentations}, last accessed 2023-06-19} The EDC consortium includes multinational companies such as Microsoft, BMW and SAP.

Following the developments of IDS and Gaia-X, there have been a number of publications that discuss use cases and challenges of dataspaces in different domains, e.\,g.\ in the context of spatial data infrastructures and INSPIRE~\citep{DBLP:journals/ijgi/KotsevMTCL20} or in the context of a data ecosystem within smart cities/municipalities~\citep{DBLP:journals/data/CunoBTLS19}.


In the following survey we discuss the architecture, data model, identity management and usage control of the individual solutions.
\paragraph{Architecture.} 
\begin{figure}
    \centering
    \includegraphics[width=0.7\linewidth]{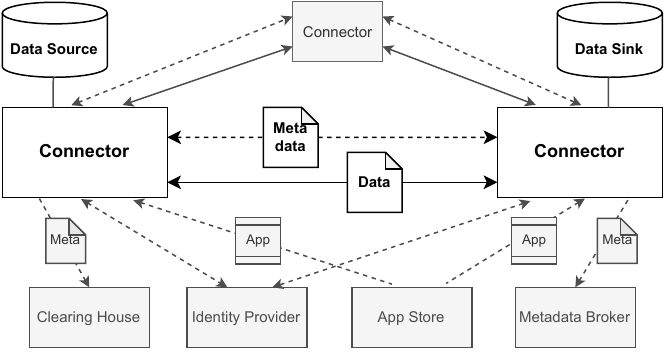}
    \caption{Interactions between the connectors and the other components in an IDS ecosystem, according to the IDS Reference Architecture \citep{idsram}.}
    \label{fig:connectorarchitecture}
\end{figure}

\Cref{fig:connectorarchitecture} shows the high-level architecture of an IDS ecosystem as described in the IDS Reference Architecture~\citep{idsram}. The central component is the connector, which allows offering data -- referred to as \textit{meta} information -- and exchanging data with other participants. The connectors integrate into each participant's infrastructure, i.\,e.\ the \textit{data source} and \textit{data sink}. Other components of such an ecosystem, which we do not discuss in detail in this paper, include the Clearing House, which serves as a logging service that records information relevant for clearing and billing as well as usage control, and the App Store, which provides apps to the connectors that perform various tasks in the IDS ecosystem (such as transforming, cleansing or analysing data). The tasks of identification, authentication, and authorization are undertaken by an Identity Provider component. According to the IDS architecture, there are central certificate authorities which are responsible for issuing and managing identity claims. We discuss the respective services in \Cref{sec:survey}.


The Metadata Broker represents a central component which registers and distributes metadata about which data offers are available from which participant and which usage conditions apply for using those data. Essentially, it fulfils the tasks of a \textit{data catalog} across the dataspace by ensuring transparency about offers and corresponding usage policies. While the IDS architecture describes a central component, there are also approaches to build a \textit{federated catalog} of data offers: Instead of directly loading/pushing the metadata into a central catalog, they are offered to other connectors (i.\,e.\ to the individual federated catalogs of the members), which requires the dataspace members to regularly monitor each other for updates regarding the offers. An example of such a federated system is the catalog envisioned in the EDC project (cf.\ \Cref{ssec:edc}).



\paragraph{Data Model.} 
The IDSA published an IDS Information Model as an RDFS/OWL ontology~\citep{OWL} that defines the types of digital content exchanged in the IDS infrastructure. The model is mainly developed by Fraunhofer FIT and Fraunhofer IAIS and maintained on GitHub; the latest version is 4.1.0, published under Apache License 2.0.\footnote{\url{https://w3id.org/idsa/core-410}, last accessed 2023-06-19} The connectors discussed in \Cref{sec:survey} all state that the respective data model is based on this information model.

\paragraph{Identity Management.}
Identity and trust mechanisms form the basis for considerations of data protection as well as access and usage rights. These mechanisms are intended to enable unique identification in a federated environment. In the context of dataspaces, trust between the participating organizations and components is established through a public key infrastructure, either through a central certification authority (cf.\ the DAPS service of the IDSA) or through the use of decentralized identifiers (cf.\ the EDC project). 


\paragraph{Usage Control.}
The usage control layer manages the rules for using and sharing the data. These rules are defined either by the owner of the data, the provider, by the government or the previous owner~\citep{DBLP:journals/cacm/PretschnerHB06}.
Data usage control can only be enforced if the respective mechanisms are installed on the data consumer side~\citep{DBLP:journals/cacm/PretschnerHB06} -- in a dataspace ecosystem this is the task of the connectors. 

Many policy languages have been suggested for usage control. The \textit{Open Digital Rights Language (ODRL)}~\citep{ODRL}, originally intended for expressing licenses, allows for simple rules and constraints expressed through logical operators. However, the ODRL and its derivatives lack guidance for specific conditions like time, location and quantity which are needed for detailed usage policies. The usage control approaches of most of the connectors discussed in this paper are based on the ODRL.

\section{Dataspace Connector Survey\label{sec:survey}}

A dataspace connector can be described as a trustworthy software component that supports the definition of usage control policies and the verifiable enforcement of those policies.
In this short survey, we select current open-source projects that develop IDS-compliant dataspace connectors. We compare these initiatives' technical details in order to provide insights into the current level of maturity and readiness.

\begin{table*}[th!]
    \begin{minipage}{\textwidth}
    \centering
    \caption{Open-source connectors that are currently available/developed.}
    \begin{tabular}{p{0.9cm}p{2.92cm}p{9.78cm}}
    \bf Name & \bf Main contributor & \bf Github project \\ \midrule
    DSC & sovity GmbH & \href{https://github.com/International-Data-Spaces-Association/DataspaceConnector}{gh:International-Data-Spaces-Association/DataspaceConnector} \\
    EDC & EDC Consortium\footnote{Consortium members include BMW, the Fraunhofer Society, Microsoft and SAP.} & \href{https://github.com/eclipse-edc/Connector}{gh:eclipse-edc/Connector} \\
    TRUE & Engineering\footnote{Engineering Ingegneria Informatica S.p.A., \url{https://www.eng.it/}} & \href{https://github.com/Engineering-Research-and-Development/true-connector}{gh:Engineering-Research-and-Development/true-connector} \\
    Trusted & Fraunhofer AISEC & \href{https://github.com/Fraunhofer-AISEC/trusted-connector}{gh:Fraunhofer-AISEC/trusted-connector} 
    \end{tabular}
    \label{tab:connectors}
    \end{minipage}
\end{table*}

\begin{table*}[th!]
    \begin{minipage}{\textwidth}
    \centering
    \caption{Key characteristics of the open-source connectors (last accessed 2023-06-19).}
    \begin{tabular}{llrrll}
    \bf Name & \bf Created & \bf Stars & \bf Commits & \bf Language & \bf Usage control \\ \midrule
    DSC & 2020-10-07 & 20\footnote{The base repository of the forked DSC currently lists 96 stars.} & 2600 & Java & IDS lang. (ODRL-based) \\
    EDC & 2021-07-26 & 167 & 1653 & Java &ODRL-based \\
    TRUE & 2020-10-30 & 18 & 82 & Java & Platoon/MyData app (IDS lang.) \\
    Trusted & 2017-09-05 & 3 & 2083 & Kotlin & LUCON lang.
    \end{tabular}
    \label{tab:connectorstats}
    \end{minipage}
\end{table*}

The selected projects are based on the connectors listed in the Data Connector Report by the \cite{idsconnectorreport}. 
We reduced the list to those that are available open-source in a public repository (e.\,g.\ on GitHub), cf.\ \Cref{tab:connectors}. Note that although some other connectors are also listed as open-source in the report, their repositories are not actually publicly available, so we did not include them in the survey; the report (published November 2022) lists 16 connectors, 9 of which as open-source, but only 4 actually provide their source in a public repository: the IDS Dataspace Connector (DSC) by sovity, the Eclipse Dataspace Connector (EDC), the TRUsted Engineering (TRUE) Connector and the Trusted Connector by Fraunhofer AISEC.

Looking at the code repositories of the connectors (cf. \Cref{tab:connectorstats}), the most popular connector in terms of GitHub stars is the EDC. The most actively developed connectors in terms of number of commits are the DSC and the Trusted Connector.\footnote{However, the last changes in the Trusted Connector repository date back to April 2022.} 

\subsection{IDS Dataspace Connector (DSC)}\label{ssec:dsc}

The DSC was initially developed by Fraunhofer ISST but is now maintained by the company sovity. The connector uses the IDS Information Model and IDS Messaging Services for message handling and provides a REST API for managing datasets by means of their metadata as IDS resources. It supports TLS-encrypted communication with other IDS connectors, data usage control rules, and integration with an identity provider. Furthermore, the developers state that it is designed to provide companies with an easy and trustworthy entry into the IDS. At the time of writing (in February 2023), the DSC is the only IDS connector that supports the enforcement of usage condition classes. The deployment of the connector can be done in Docker or Kubernetes.

\paragraph{Architecture.}
The architecture and functionalities of the IDS Connector are defined by the IDS Reference Architecture Model (RAM)~\citep{idsram}. The architecture of the connector potentially consists of several components, including the core connector at its centre (responsible for the identity management, usage control, communication, etc.), the data management component (which controls the data sources, endpoints and contracts) and optionally a GUI that offers a dashboard, catalog and policy templates.\footnote{An architecture diagram can be found in the online documentation of the IDS Connector: \url{https://international-data-spaces-association.github.io/DataspaceConnector/Documentation/v6/Architecture}, last accessed 2023-06-19} 
The external IDS Metadata Broker, which allows to (un)register and query data offers of the dataspace members, is reached by the connector via the defined messaging services.


\paragraph{Data Model.} 
The data model of the IDS Connector\footnote{\url{https://international-data-spaces-association.github.io/DataspaceConnector/Documentation/v5/DataModel}, last accessed 2023-06-19} is based on the IDS Information Model~\citep{Lange2022TheExchange}. The metadata of a dataset/data source is called a \textit{resource} and includes e.\,g.\ the title, a description and license information. The resources of an IDS Connector are organized by \textit{catalogs}. Additionally, a resource can hold a list of \textit{contract offers}. Contract offers describe the potential consumer and provider of the data and can consist of multiple rules, so-called IDS Usage Control Patterns (see below), which control the access and usage. In case the provider and data consumer agree on the usage conditions, this is stated in a \textit{contract agreement}. \Cref{lst:idsresource} displays an example resource which is listed in a catalog. The example resource specifies a permission (as a rule in the attached contract offer).

\paragraph{Identity Management.}
To establish ``trust'' between the participating IDS connectors, a public key infrastructure through a central certification authority is used. The individual identity certificates of the participating components are managed by a central authority, the \textit{Dynamic Attribute Provisioning Service} (DAPS). Connectors request a digitally signed JSON web token (JWT) from the DAPS in order to authenticate themselves. The DSC communicates with the DAPS provided by Fraunhofer AISEC by default. It is available at \texttt{https://daps.aisec.fraunhofer.de/}. The implementation of DAPS is open-source and can be accessed on GitHub.\footnote{\url{https://github.com/International-Data-Spaces-Association/IDS-G}, last accessed 2023-06-19}

\paragraph{Usage Control.}
The DSC supports usage policies written in the IDS Usage Control Language, which is based on the ODRL~\citep{ODRL}. It enables so-called ``IDS Contracts'', which are divided into two main sections: contract-specific metadata and the IDS Usage Control Policy. The policy defines several Data Usage Control statements called \textit{IDS Rules} that are specified in the IDS Usage Control Language and consist of permissions, prohibitions and obligations. The technically enforceable rules are then transformed to a technology-dependent policy for Usage Control enforcement. The DSC currently implements 9 out of the 21 policy classes defined by the IDSA Position Paper on Usage Control~\citep{usagecontrolreport}. \Cref{lst:idsresource} holds an example policy that permits use, i.\,e.\ gives access, for the stated validity period of the contract offer.

\subsection{Eclipse Dataspace Connector (EDC)\label{ssec:edc}}
The EDC\footnote{\url{https://github.com/eclipse-edc/Connector}, last accessed 2023-06-19} is a framework that implements the IDS standard as well as part of other services in the Eclipse Dataspace Components required to build a dataspace. 
The project has a documentation website;\footnote{\label{fn:edc-docs}\url{https://eclipse-edc.github.io/docs}, last accessed 2023-06-19} however, the majority of relevant information is instead found in Markdown files inside the different repositories.

\begin{figure}
    \begin{minted}[frame=single,fontsize=\scriptsize]{sparql}
@prefix ids: <https://w3id.org/idsa/core/> .
@prefix xsd: <http://www.w3.org/2001/XMLSchema#> .
<https://ex.org/api/catalogs/cat1>
  a <https://w3id.org/idsa/core/ResourceCatalog> ;
  ids:offeredResource <https://ex.org/api/offers/res1> .
<https://ex.org/api/offers/res1>
  a ids:Resource ;
  ids:contractOffer <https://ex.org/api/contracts/cont1> ;
  ids:description "Resource of an IDS Connector" ;
  ids:title "Example Resource" .
<https://ex.org/api/contracts/cont1>
  a ids:ContractOffer ;
  ids:contractEnd "2022-12-24"^^xsd:dateTimeStamp ;
  ids:contractStart "2022-12-25"^^xsd:dateTimeStamp ;
  ids:permission <https://ex.org/api/rules/policy1> .
<https://ex.org/api/rules/policy1>
  a ids:Permission ;
  ids:action <https://w3id.org/idsa/code/USE> ;
  ids:description "provide-access" ;
  ids:title "Example Usage Policy" .
    \end{minted}
    \caption{Example of an IDS catalog, resource, contract offer and permission in RDF Turtle syntax. The example uses the IDS Information Model~\citep{Lange2022TheExchange}.}
    \label{lst:idsresource}
\end{figure}

\paragraph{Architecture}

The EDC is designed to align with the standards and guidelines established by Gaia-X. It supports IDS-based messages and policy definitions, which are also used by Gaia-X, and participates in IDSA committees and working groups to pursue its alignment with both Gaia-X and the IDSA\footnote{\label{fn:edc-pptx}\url{https://github.com/eclipse-edc/Collateral/blob/main/Latest\%20Presentations/2022-04-26\%20Eclipse\%20Dataspace\%20Connector\%20-\%20Overview\%20Deck.pdf}, last accessed 2023-06-19}.
The EDC consists of a Connector, a Federated Catalog Node, a Federated Catalog Crawler, a Policy Management, a Data Asset Management, an Identity Hub, and a Registration Service\footref{fn:edc-pptx}.


The goal of the project is to work on a decentralized data space implementation: while, according to the documentation, the EDC shall support IDSA-based components like the DAPS service for identity management and the Metadata Broker, it aims to implement decentralized approaches such as identity management via Decentralized Identifiers~\cite{DID} or Federated Catalogs.

A Federated Catalog provides data contract offers while also collecting offers from other dataspace participants. The identity information of a participant is provided by their Identity Hub, while the Registration Service contains information about all participants of a dataspace. 
While earlier presentations\footref{fn:edc-pptx} mentioned Policy Management as well as Data Asset Management components, there is currently no detailed description provided at that time of writing.\footref{fn:edc-docs}



\paragraph{Data Model.}

The EDC is based on IDS standards and uses terms from the IDS Information Model. In earlier versions and presentations\footref{fn:edc-pptx} it differed in some details, e.\,g.\ shared resources are called assets, which are symbolic links to datasets; contracts defined the data exchange between two parties, with different stages (starting with ``contract definition''); contract offers are instantiations of a contract definition for a specific consumer. 

\paragraph{Identity Management.}
The identity management of the EDC is explained in an earlier presentation:\footref{fn:edc-pptx} an EDC dataspace can either have a central dataspace authority or a fully decentralized one. A connector needs a decentralized identifier (DID)~\citep{DID} anchored in a trust framework, as well as a self-description that consists of attributes of the member, a list of verifiable credentials (VC)~\citep{VC}, a pointer to the identity hub, a pointer to claims and proofs, and URLs of required endpoints. In case of a central dataspace authority, a participant is verified by querying the central authority and verifying the membership as well as the claims using the DID and VC. In case of a fully decentralized approach, the VC need to be signed by other members and are replicated among them along with the member registry and the policies.

\paragraph{Usage Control.}
The EDC aims to supports usage and access policies which are based on the ODRL policy model. In a recent sample code,\footnote{\url{https://github.com/eclipse-edc/Samples/blob/main/transfer/transfer-07-provider-push-http/README.md\#3-create-a-policy-on-the-provider}, last accessed 2023-06-19} the connector uses an ODRL-compliant policy (in JSON-LD syntax); however, at the time of writing, the sample policy is very simplistic (no specific constraints, allowing ``direct access to all assets'').

 
The idea of policies in the EDC is aligned with ODRL. Policies are sets of rules specifying allowed, disallowed or required actions of a consumer for requested assets: Permissions define allowed actions on the asset and can include conditions. Prohibitions disallow specific actions, while obligations require the consumer to perform certain actions.\footnote{\url{https://github.com/eclipse-edc/Connector/blob/main/docs/developer/architecture/usage-control/policies.md}, last accessed 2023-06-19} 

\Cref{lst:edcpolicy} demonstrates the implementation of policies in the EDC program code.

\begin{figure}
    \begin{minted}[frame=single,fontsize=\scriptsize]{java}
var readPermission = Permission.Builder.newInstance().action(Action.Builder
    .newInstance().type("READ").build()).build();
var distributeProhibition = Prohibition.Builder.newInstance().action(Action.Builder
    .newInstance().type("DISTRIBUTE").build()).build();
var readNotDistributePolicy = Policy.Builder
    .newInstance().id("use-all").permission(readPermission)
    .prohibition(distributeProhibition).target("test-document").build();
    \end{minted}
    \caption{Example Java code of an EDC policy for allowing read access and disallowing further distribution of an asset.}
    \label{lst:edcpolicy}
\end{figure}





\subsection{TRUsted Engineering (TRUE) Connector}

The TRUE Connector is an open-source connector that enables trusted data exchange and allows participation in an IDS ecosystem, as well as governance models, to facilitate secure and standardized data exchange and linkage. According to the connector's documentation, it is compliant with the latest IDS specifications.

\paragraph{Architecture.}
The architecture of the TRUE Connector consists of three components: The Execution Core Container (ECC) is responsible for the data exchange within the IDS ecosystem, representing data using the IDS Information Model, interacting with external Identity Providers and communicating with an IDS Broker for registration and querying information. The Back-End (BE) Data Application represents a somewhat trivial data application for generating and consuming data on top of the ECC component. The last component is the Usage-Control (UC) Data Application, which, however, does \emph{not} include contract negotiation procedures. The first two are licensed under AGPLv3, while the UC App's license is still to be determined.

\paragraph{Data Model.}
The ECC data model is based on the IDS Information Model. 

\paragraph{Identity Management.}
Identity provider support is off by default, but can be turned on in the settings. 
The TRUE Connector supports identity verification using the IDS Dynamic Attribute Provisioning Service (DAPS). 

\paragraph{Usage Control.}
Usage control is optional and turned off by default, but turning it on is mandatory for contract negotiation.
The TRUE Connector supports both, the Platoon Usage Control Data App and the MyData Usage Control Data App; the default is Platoon. The Platoon Usage Control module is a modification of the IDS Dataspace Connector (DSC) and supports usage policies written in the IDS Usage Control Language. It includes REST services for getting, uploading, and removing Contract Agreements. Additionally, it includes a REST service for applying usage control enforcement on input data according to the related Contract Agreements. \Cref{lst:truepolicy} displays a sample Contract Agreement in the IDS Usage Control Language.

\begin{figure}
    \begin{minted}[frame=single,fontsize=\scriptsize]{json}
{"@context": 
  {"ids": "https://w3id.org/idsa/core/", "idsc": "https://w3id.org/idsa/code/"},
  "@type": "ids:ContractAgreement",
  "@id": "https://ex.org/contract/restrict-access-interval",
  "ids:permission": [{
      "ids:action": [{"@id": "idsc:USE"}],
      "ids:constraint": [{
          "@type": "ids:Constraint",
          "ids:leftOperand": {"@id": "idsc:POLICY_EVALUATION_TIME"},
          "ids:operator": {"@id": "idsc:TEMPORAL_EQUALS"},
          "ids:rightOperand": {
            "@type": "ids:interval",
            "@value": {
              "ids:begin": 
              {"@value": "2021-06-15T00:00:00Z", "@type": "xsd:datetimeStamp"},
              "ids:end": 
              {"@value": "2021-12-31T00:00:00Z", "@type": "xsd:datetimeStamp"}
            }}}]}]}
    \end{minted}
    \caption{Example of a Contract Agreement in the MyData Usage Control Data App, written in the IDS Usage Control Language. The contract restricts the use by a given time interval.}
    \label{lst:truepolicy}
\end{figure}

\subsection{Trusted Connector}

The IDS Trusted Connector is an open-source framework developed by Fraunhofer AISEC. It describes itself as an open IoT edge gateway platform and states that it implements the IDS Reference Architecture\footnote{\url{https://industrial-data-space.github.io/trusted-connector-documentation/docs/overview/}, last accessed 2023-06-19} and that it offers a range of protocol adaptors to connect sensors with cloud services and other Connectors. 

\paragraph{Architecture}
The Trusted Connector consists of a Core Container and one or more Application Containers. The Core Container provides the main functionality and is the sole intermediate between the Application Containers and the internet. It is based on Spring Boot\footnote{\url{https://spring.io/projects/spring-boot}, last accessed 2023-06-19} and provides a REST API. Services include Usage Control, Route Manager, and the IDS protocol. The Route Manager allows for message routing based on Apache Camel, and includes protocol adapters for secure communication between Trusted Connectors. Applications come in the form of Docker containers, and are initially isolated from each other and restricted in a virtual network with the Core Container. This prevents malicious applications from interfering with the running system.

\paragraph{Identity Management}
Trusted Connectors use the IDS Dynamic Attribute Provisioning Service (DAPS) to authorize each other in order to establish cross-enterprise authorization. Every connector possesses a unique identifier embedded in a X.509 device certificate which is used to authenticate the connector instance. A Dynamic Attribute Token (OAuth Access Token) is used to store attributes such as certification status or IDS membership. 
To perform this authorization, the requesting connector needs to present its device certificate to the DAPS in order to receive a Dynamic Attribute Token. This token is then used to access the connector's resources.

\paragraph{Usage Control}

\begin{figure}
    \begin{minted}[frame=single,fontsize=\scriptsize]{scheme}
flow_rule {
  id anonymized                                     // Rule id
  description "Don't leak personal or internal"
  when publicEndpoint                               // Target identifier
  receives { label(personal) or label(internal) }   // Received message labels
  decide drop }                                     // Drop message
    \end{minted}
    \caption{Example of a LUCON policy to anonymize personal data before they are transmitted.}
    \label{lst:luconpolicy}
\end{figure}

The Trusted Connector provides a policy framework for ``Logic Based Usage Control'' to enable the usage control of data flows between Apps and Connector instances. \textit{LUCON}, the policy language for controlling data flows, allows the labeling of data and the enforcement of specific actions to be carried out. Data can be limited from leaving a connector, anonymized, or logged. For example, a policy may state that all personal data must be anonymized before it leaves the Connector, or that data must be deleted after 30 days. 
A LUCON policy consists of two parts: a \textit{flow rule} and a \textit{service description}. The flow rule defines the conditions under which the policy should be applied, and the service description defines the services to which the rule applies.
\Cref{lst:luconpolicy} displays an example flow rule to anonymize personal data before leaving the connector: the rule \textit{anonymized} declares that if a service matching the \texttt{publicEndpoint} description receives a message that contains the label ``personal'' or ``internal'', the message has to be dropped.


\section{Conclusion\label{sec:conclusion}}

The adoption and use of dataspaces in real-world use cases depends on the maturity of the available technologies. This means that dataspaces will only become successful if there are open-source connectors that are simple, well-documented and offer a high level of technological readiness. 
To assess this readiness, we have surveyed existing open-source implementations of IDS-compliant connectors with regard to their architecture, underlying data model and usage control language. In conclusion, our key findings can be summarized as follows:
\begin{compactitem}
    \item The reviewed connectors all state that they implement the IDS architecture and that they are aligned in terms of central identity management. However, the EDC framework is also developing an alternative decentralized identity verification system.
    \item We have found different approaches to set policies in the connector implementations. The implementations use ODRL and partly define their own usage control languages (e.\,g.\ the IDS Usage Control Language or the LUCON policy language by Fraunhofer AISEC). These differences pose risks of incompatibility when describing usage policies and could potentially complicate building an ecosystem across connectors.
    \item Conducting the research for this survey was complicated partly based on the software projects' documentations. We point out the surprisingly bad state of documentations and instructions of some of the projects, which pose a clear risk of limited up-take and dissemination of the tools. 
    Given the state of documentations, we have noted a rather low level of technical readiness/maturity, which in its current state additionally hinders the adoption of connectors in industry.
\end{compactitem}

An important aspect that we will address in future work is a more structured benchmarking of the existing implementations by setting up defined tasks and experiments to assess interoperability between systems, maturity, as well as performance and scalability. 


\bibliography{literature}

\end{document}